\documentclass[useAMS,usenatbib, usegraphicx]{mn2e}
\usepackage{lscape}
\usepackage{rotating}
\usepackage{subfigure}

\def \kms{\ifmmode{~{\rm km\,s}^{-1}}\else{~km~s$^{-1}$}\fi}
\def \vhel{\ifmmode{V_{{\rm hel}}}\else{$V_{{\rm hel}}$}\fi}
\def \vsys{\ifmmode{V_{{\rm sys}}}\else{$V_{{\rm sys}}$}\fi}
\def \degree{\ifmmode{^{\circ}}\else{$^{\circ}$}\fi}

\def \myr{\ifmmode{{\rm\ M}_\odot{\rm\ yr}^{-1}}\else{${\rm\ M}_\odot$ 
yr$^{-1}$}\fi}

\def \mdot{\ifmmode{{\rm\dot{M}}}\else{${\rm\dot{M}}$}\fi}

\newcommand{\OIII}{[O\,{\sc iii}]\ $\lambda$5007\,\AA}

\newcommand{\NII}{[N\,{\sc ii}]\ $\lambda$6584\,\AA}

\def \apj{Astrophysical Journal}
\def \mnras{Monthly Notices of the Royal Astronomical Society}

\def \aap{Astronomy and Astrophysics}
\def \aj{Astronomical Journal}

\def \apjs{The Astrophysical Journal Supplement Series}
\def \apjl{The Astrophysical Journal Letters}

\title[The kinematics of the large western knot in NGC~6543]{The kinematics of
 the large western knot in the halo of the young planetary nebula NGC~6543}
\title[The kinematics of the large western knot in NGC~6543]{The kinematics of
 the large western knot in the halo of the young planetary nebula NGC~6543}
\author[Deborah L. Mitchell et al.]{Deborah L. Mitchell$^{1}$,\thanks{E-mail:
dlm@jb.man.ac.uk (DLM)} M. Bryce$^{1}$, J. Meaburn$^{2}$, J.A. L\'{o}pez$^{2}$
, \newauthor M. P. Redman$^{3}$, D. Harman$^{3}$,
 M. G. Richer$^{2}$
 and H. Riesgo$^{2}$\\
$^{1}$Jodrell Bank Observatory, University of Manchester, Macclesfield
SK11 9DL, UK
 \\
$^{2}$Instituto de Astronom{\'\i}a, Universidad Nacional Autónoma de M\'{e}xico,
 Apartado Postal 877, 22800 Ensenada, B.C., M\'{e}xico
 \\
$^{3}$Department of Physics, National University of Ireland Galway, College Rd., Galway, Ireland.
\\
$^{4}$Astrophysics Research Institute, Liverpool John Moores University, Twelve Quays House, Egerton Wharf,
 CH41 1LD, UK
}
\begin{document}

\date{Accepted ... Received...}

\pagerange{\pageref{firstpage}--\pageref{lastpage}} \pubyear{2002}

\maketitle

\label{firstpage}

\begin{abstract}
A detailed analysis is presented of the dominant ionised knot in the halo of
 the planetary nebula NGC~6543. Observations were made at high spectral and 
spatial resolution of the \OIII\ line using the Manchester echelle 
spectrometer combined with the 2.1-m San Pedro Martir Telescope. A 20-element 
multislit was stepped across the field to give almost complete spatial 
coverage of the large western knot and surrounding halo. 

The spectra reveal, for the first time, gas flows around the kinematically 
inert knot. The gas flows are found to have velocities comparable to the 
sound speed as gas is photo-evaporated off an ionised surface. No evidence is 
found of fast wind interaction with the knot and we find it likely that the 
fast wind is still contained in a pressure-driven bubble in the core of the 
nebula. This rules out the possibility of the knot having its origin in 
instabilities at the interface of the fast and AGB winds. We suggest that the 
knot is embedded in the slowly expanding Red Giant wind and that its surfaces 
are being continually photoionised by the central star.
\end{abstract}


\section{Introduction}

NGC~6543 (RA $17^{\mathrm{h}}\,58^{\mathrm{m}}\,33\fs4$\ \&\ Dec$
+66^{\circ}\,37\arcmin\,59\farcs0$ [J2000.0]) is one of the brightest and 
most structurally complex planetary nebulae (PNe) known. The central star is 
an 07+WR type, which has a high-speed (1900 kms$^{-1}$) particle wind 
(Patriarchi \& Perinotto 1991). The bright core of the nebula appears as two 
concentric crossed ellipses in projection with their major axes orientated at 
position angles  $25^{\circ}$ and $115^{\circ}$ respectively (Miranda \& Solf 
1992). \cite{1987AJ.....94..948B} detected two bright polar caps aligned with 
the major axis of the $25^{\circ}$ ellipse and two twisted jet-like features 
extending radially outwards from the caps. \cite{2001AJ....121..354B} 
discovered a series of nine regularly spaced concentric rings around the core 
of NGC~6543 in archival HST images. The best estimate of distance to NGC~6543 
is 1001 $\pm$ 269 pc (Reed et al. 1999), which was measured directly from the 
expansion parallax.

 \begin{figure*}
\centering
\mbox{\resizebox{12.0cm}{!}{\includegraphics{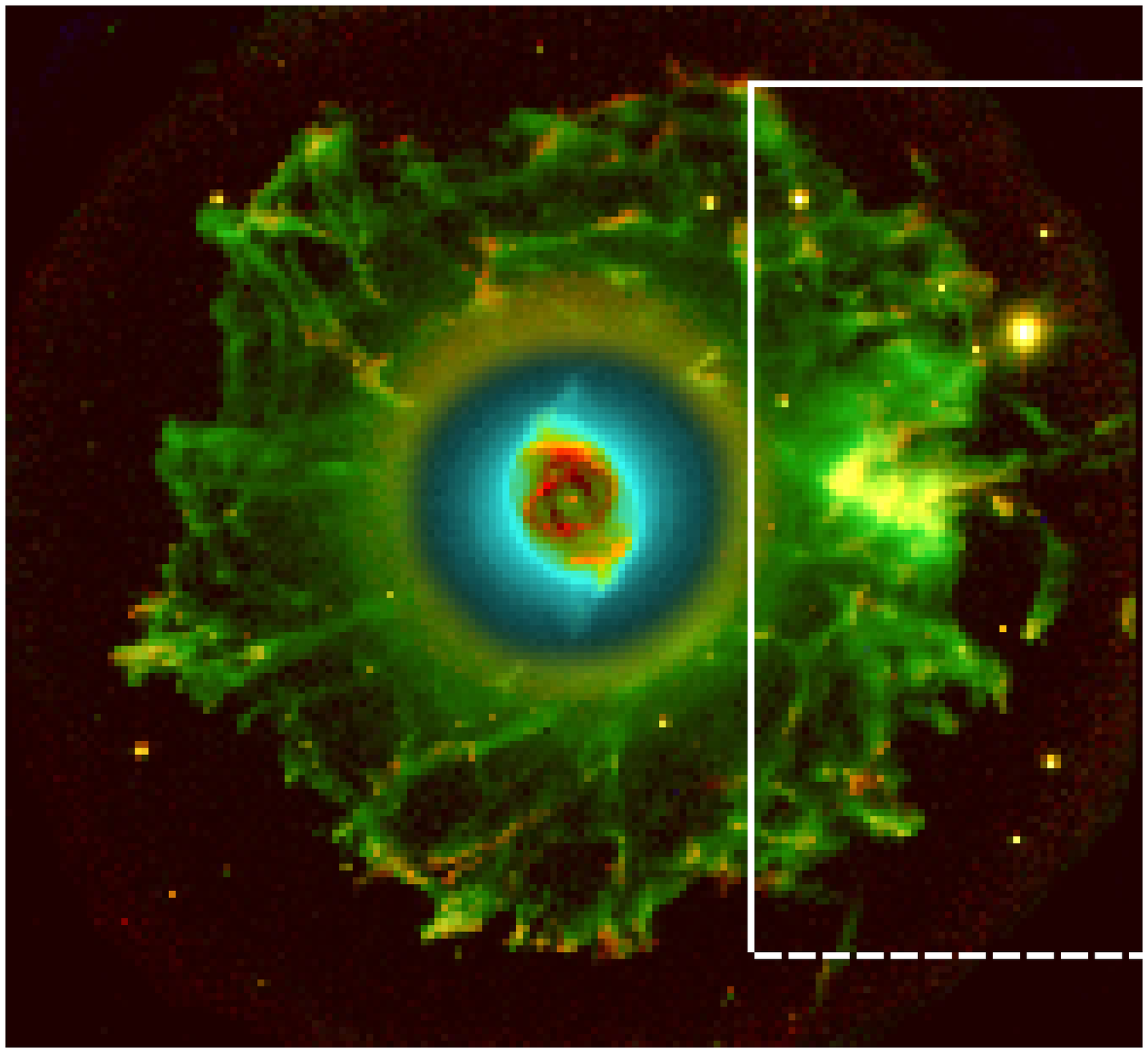}}}
\caption {A composite image showing the core and halo of NGC~6543 taken with 
the Nordic Optical Telescope. \NII\ is shown in red and \OIII\ in blue and 
green. The image is 385\arcmin\ by 352\arcmin\ in extent. North is to the top 
and east is to the left. The dashed box outlines the region covered by the 
multislit (see section 2). The image shows the complex nebula core, which 
consists of two crossed ellipses in projection, and jets protruding from the 
polar caps of the shell. Also visible is the large, filamentary halo 
surrounding the core and the prominent knot in the west. Credit R. Corradi 
(Isaac Newton Group) and D. Goncalves (Inst. Astrofiscia de Canaries).}
\end{figure*}

The bright core of NGC~6543 is surrounded by a faint, extended,
spherical halo (Millikan 1974) of radius 165 arcsec (Middlemass, Clegg,
\& Walsh 1989). \cite{1987ApJS...64..529C} defined NGC 6543 as having
a Type I detached halo.  The halo has a very filamentary appearance and 
contains a large, bright,
irregularly-shaped knot located 100 arcsec west of the
core. The halo of NGC~6543 is unsually bright in \OIII\ (Middlemass et al. 
1989) despite the fact that the main PN shell also shows bright
\NII\ emission, suggesting a complex overall
structure. \cite{1992MNRAS.254..477B} found the halo to be
kinematically inert with a maximum expansion velocity of 4.5
kms$^{-1}$, despite its explosive, filamentary appearance. They found
that both the bright knot and halo are centred on the systemic heliocentric 
radial velocity of \vsys = $-68.5$ kms$^{-1}$, confirming that the knot and
halo are physically linked. An \OIII\ and \NII\ composite image that features 
both the halo and core of NGC~6543 is shown in Fig.~1 with the dominant 
ionised knot being considered here.
 
The bright knot in the halo of NGC~6543 is highly unusual in that it is 
relatively large and is a unique feature within the halo.  
\cite{1989MNRAS.239....1M} calculated that 
the knot has an electron temperature of $\mathrm{T}_{\mathrm{e}}$ = 14700~K 
using 
[O\,{\sc iii}]$\lambda\lambda$4959\,+\,5007/4363 line intensity ratios. 
\cite{1991MNRAS.252..535M} derived a much lower $\mathrm{T}_{\mathrm{e}}$ of 
8860$\pm$1000 K from observations of the thermal broadening of the 
H$\alpha$ and \NII\ lines. In order to account for this discrepency, a 
mass-loaded wind model was proposed, which predicts that a supersonic stellar 
wind percolates through the clumpy core and mass-loads in the process. The 
mass-loaded wind then percolates into the halo, which adds heat and raises the 
temperature of the \OIII\ region (but see section 4.1 for an alternative 
explanation).

At present, very little is known about the origin of the knot and what 
physical processes have affected its morphology since its formation. The low 
global expansion velocity of the halo, the low turbulence in the bright knot 
gas and the evidence that an ionisation front is present on the knot surface 
(the \NII\ ridge is displaced from the parallel \OIII\ one) all suggest that 
the large knot is the dense relic of the Red Giant wind, now being 
photoionised on its surface. The observations to be reported here aim to 
determine whether or not this knot, and hence the whole halo, is subject to 
the fast wind from the central star; if so, high-speed flows of ablated gas 
would be expected in the faint regions in the close vicinity of this knot. 
For this purpose, \OIII\ line profiles have been obtained with a stepped 
multi-slit, giving unprecendented spatial coverage, over the whole vicinity 
of the unusual knot.

\section[]{Observations and Data Reduction}

The Manchester echelle spectrometer (Meaburn et al. 2003) combined with 
the 2.1-m, f/8 San Pedro Martir (MES-SPM) telescope was used in its secondary 
mode to take direct images at \OIII\ and \NII\ of the bright knot located in 
the western part of the halo of NGC~6543. The images, shown in Figs. 2 and 3, 
were taken during an observing run on the SPM telescope in Mexico in June 
2004, using a SITe3 CCD with 1024$\times$1024 24 $\mu$m square pixels 
($\equiv$ 0.31\arcsec pixel$^{-1}$).  Integration times were 1800s for both 
images. 

Spatially resolved, longslit observations of the bright knot were taken using 
the MES in its primary spectral mode during the same observing run. The 
observations were taken through a narrow-band (30 \AA) filter, which isolates 
the \OIII\ emission line in the 114th echelle order.

A 20-element multislit was used to cover the knot with parallel north-south 
slit positions. Each slit is separated by 10.75 arcsec ($\equiv$ 69
kms$^{-1}$), each slit being 70 $\mu$m wide ($\equiv$ 0.9\arcsec\ and 5
kms$^{-1}$). Three integrations of 1800~secs were obtained, between each of 
which, the multislit was offset to the west by approximately 3 arcsec. The 
net result is almost total spatial coverage of the knot by 60 parallel 
north-south slits, as shown in Fig. 4. 2$\times $2 binning was 
adopted during the observations, giving 512 pixels in the spectral (x) 
direction ($\equiv$ 4.79 kms$^{-1}$ pixel$^{-1}$) and 512 pixels in the 
spatial (y) direction ($\equiv$ 0.62\arcsec pixel$^{-1}$). Each slit is 317 
arcsec long and a total distance of 213 arcsec in the E-W direction is 
covered by the dataset (slit 1 to slit 60).

\begin{figure*}
\centering
\mbox{\resizebox{13.0cm}{!}{\includegraphics{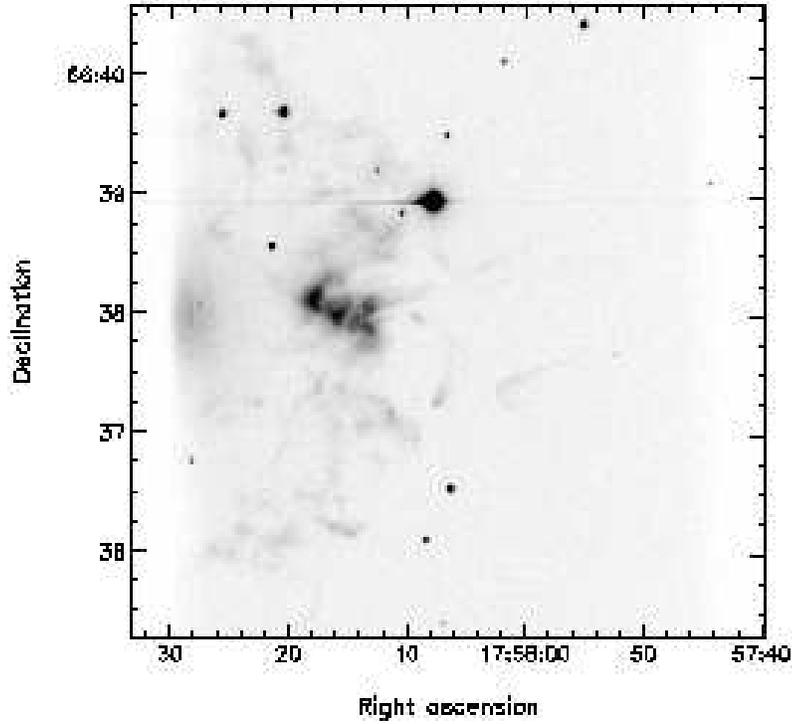}}}
\caption {A narrowband \OIII\ image of the western part of the halo of 
NGC~6543, showing the bright knot around RA 
$17^{\mathrm{h}}\,58^{\mathrm{m}}\,16\fs0$\ 
\&\ Dec$+66^{\circ}\,37\arcmin\,59\farcs0$ [J2000.0]. The central star of the 
PN is located 100\arcsec\ east of the knot. The star at 
RA $17^{\mathrm{h}}\,58^{\mathrm{m}}\,08\fs0$\ 
\&\ Dec$+66^{\circ}\,38\arcmin\,55\farcs0$ is overexposed, which has resulted 
in horizontal bleeding across the CCD. }
\end{figure*}

\begin{figure*}
\centering
\mbox{\resizebox{14.5cm}{!}{\includegraphics{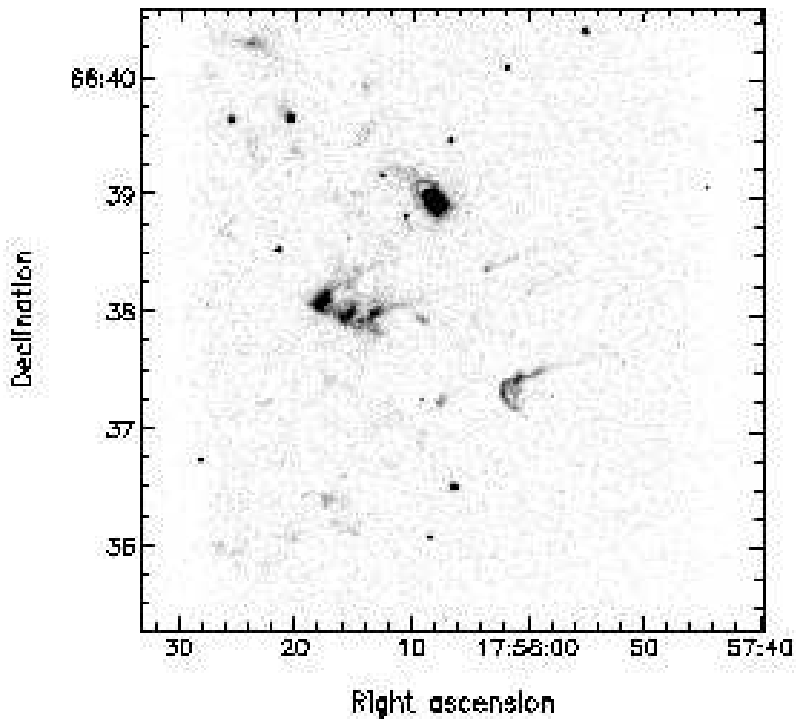}}}
\caption {A narrowband \NII\ image of the region shown in Fig. 2 (the bright 
star is contaminated by a filter ghost).}
\end{figure*}

\begin{figure*}
\centering
\mbox{\resizebox{\textwidth}{!}{\includegraphics{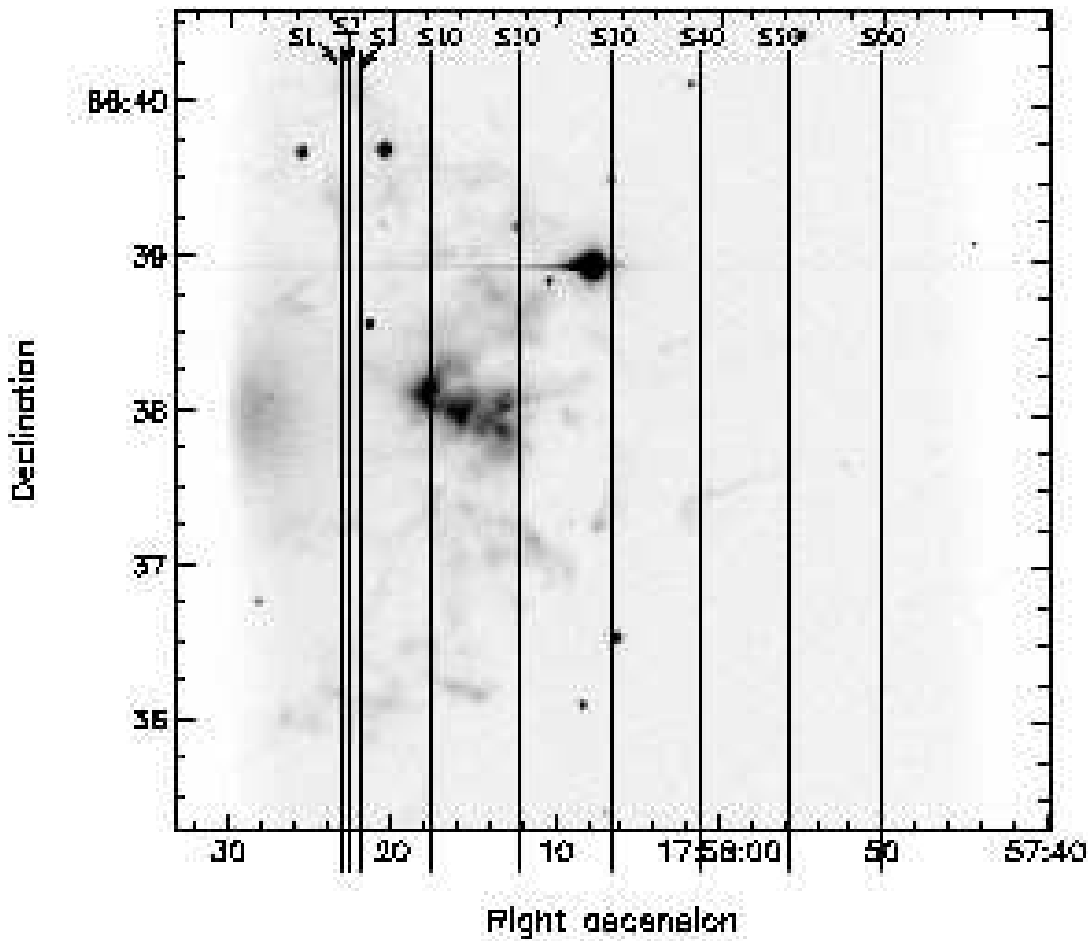}}}
\caption{\OIII\ image of the knot region (Fig. 2) with a sample of the 60 slit 
positions superimposed (S1, S2, S3, S10, S20, S30, S40, S50 and S60). Each 
slit position is separated by $\sim 3\arcsec$. The full extents of the slit 
lengths on the sky are illustrated here.}
\end{figure*}

Data reduction was performed using \textsc{starlink} software. The \OIII\
spectra were bias-corrected and cleaned of cosmic rays. The spectra were then 
wavelength calibrated against a ThAr emission line lamp. A ThAr spectrum from a single slit was used to aid the line identification process in the multislit 
spectrum due to the complication of overlapping arc lines from adjacent slits.

The three calibrated multislit \OIII\ position-velocity (pv) arrays were 
sliced up in a direction parallel to the slit lengths to give 60 pv arrays, 
each containing emission from a single slit. The 60 pv arrays
were then stacked in sequence from east to west using the program 
\textsc{formcube}
(Malcolm Currie, private communication) to produce a pv data cube where the 
E-W spatial direction has been reconstructed. The data cube has
spatial dimensions of 317\arcsec $\times$ 213\arcsec\ (in the N-S and E-W 
directions, respectively) and a dispersion dimension ranging from 
heliocentric radial velocities \vhel\ = $-34$ to $-103$ kms$^{-1}$ with a 
velocity resolution of 7 kms$^{-1}$. 

\section[]{Results}

\subsection[]{Direct images}

Fig.~5 shows a subset of the \OIII\ image (Fig.~2) in order to reveal the 
substructure in the bright knot (Meaburn et al. 1991). It is clear that
the knot is composed of several smaller sub-knots, which have been labelled 
A-F in Fig.~5 (following Bryce 1992). The sub-knots are embedded in a 
relatively bright region of gas identified in Fig~5. as `knot gas'. The 
outline of the knot gas is marked on the image.

\begin{figure*}
\centering
\mbox{\resizebox{5.2cm}{!}{\includegraphics{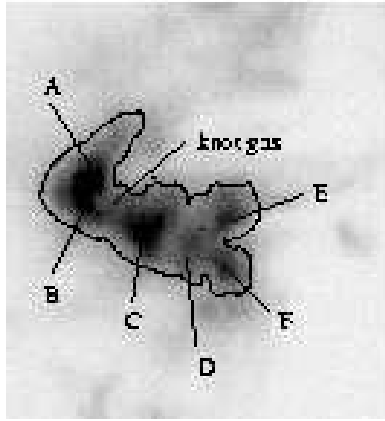}}}
\caption {An extracted region of the \OIII\ image that shows the knot and the 
faint halo immediately surrounding it.  The image is shown at high-contrast to
 reveal substructure in the knot. The brightest features (sub-knots) have been labelled 
A-F. The sub-knots are embedded in bright `knot-gas' and the boundary between 
the knot gas and the faint, background halo is outlined on the image.}
\end{figure*} 
     
\subsection[]{Combined image}

The datacube was collapsed along the dispersion axis to produce an image 
showing the whole range in \vhel. This effectively represents a 
narrowband \OIII\ image of the knot. 
 
\begin{figure*}
\centering
\mbox{\resizebox{12.0cm}{!}{\includegraphics{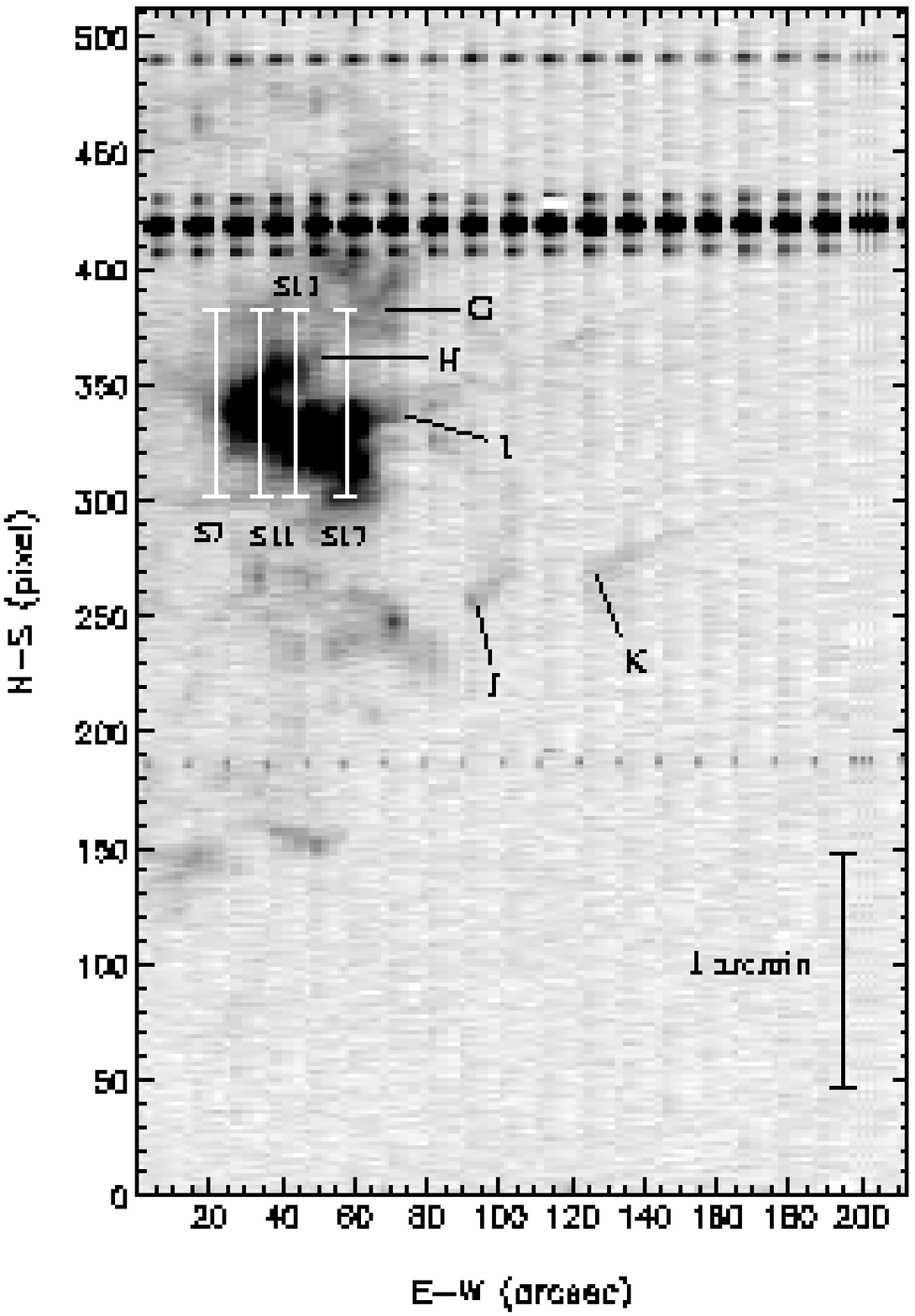}}}
\caption {The total velocity range of the \OIII\ datacube is shown at 
low-contrast in order to highlight faint structure in the halo. The N-S extent
 of the figure is shown in Fig.~1. The slit length of 512 pixels $\equiv$ 317 
arcsec. The plume of gas located to the north of the knot has been labelled G;
 two trails of gas are labelled H and I; the comet-like features in the 
south-west are labelled J and K. The lines labelled S7, S11, S13 and S17 mark 
the lengths of slits from which the corresponding \vhel\ and line width plots are 
shown in Figs.~9 and 10, respectively. The striped horizonal bands at 
cross-sections 185, 420 and 490 are continuum features from field stars (if a 
star hits one slit it appears in this image as though it is in all 20 slits).}
\end{figure*}

Fig.~6 shows this combined image at low contrast in order to highlight 
structure in the faint halo that surrounds the knot. The dark, horizontal 
bands at cross-sections 185, 420 and 490 are continuum emission from field 
stars. The image reveals that the bright knot gas is surrounded by much 
fainter `halo gas' and the boundary between the two regions is distinct. The 
faint halo gas is present both to the north and south of the knot region. The 
southern emission (cross-sections 120 to 280) has a more filamentary 
appearance than the faint halo gas in the north. A faint plume of gas is 
located north of the knot, between cross-sections 360 and 400, labelled G, 
which is orientated in a north-west direction. Trails of gas appear to emanate
 from the northern edge of the knot, labelled H and I, which have similar 
orientations to plume G. Two distinct features lie south-west of the knot, 
labelled J and K, both of which have a comet-like appearance (these features 
are also clear in Figs.~2 and 3).  

\subsection[]{Velocity Slices}

The datacube was sliced up along
the dispersion axis to create radial velocity slices, each of width 10 
kms$^{-1}$. The velocity planes within each slice were co-added to create 
two-dimensional images. The images reveal how the morphology of the knot 
varies with \vhel. Emission was detected in three of the images, $I_{1}$, 
$I_{2}$ and $I_{3}$, which are shown at high and low contrast in Figs.~7 and 
8, respectively. The high-contrast images have been normalised with respect to
 the maximum intensity. Co-ordinates of features within the images are expressed as (NS, EW) pixels and arcsecs, respectively. The NS axis is expressed in pixels for ease of comparison between Figs. 7, 8, 9 and 10. The velocity ranges covered by each image are listed in Table~1. Note that $I_{2}$ is centred on \vsys. The images show the following features and trends:

\begin{landscape}
\begin{figure}
\centering
\subfigure[$I_{1}$ (\vhel\ = $-83$ kms$^{-1}$ to 
$-73$ kms$^{-1}$)]{\mbox{\resizebox{7.0cm}{!}{\includegraphics{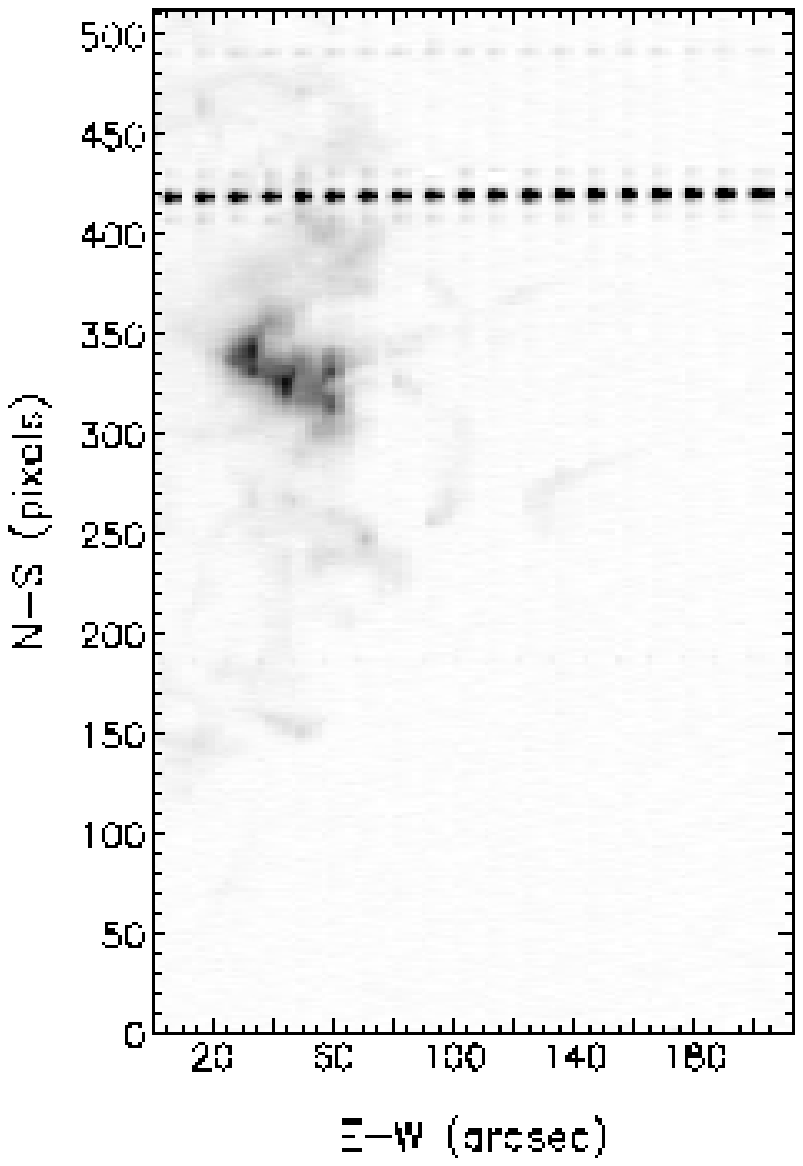}}}}
\subfigure[$I_{2}$ (\vhel\ = $-73$ kms$^{-1}$ to 
$-63$ kms$^{-1}$)]{\mbox{\resizebox{7.0cm}{!}{\includegraphics{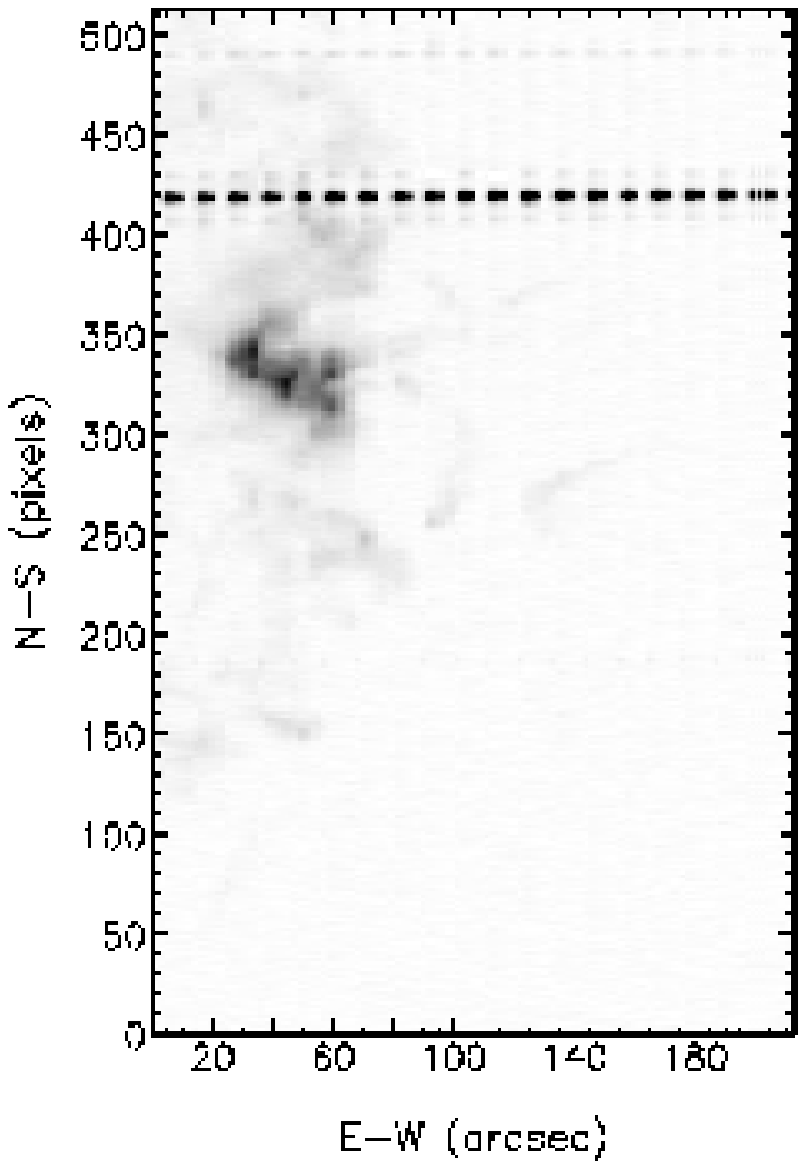}}}}
\subfigure[$I_{3}$ (\vhel\ = $-63$ kms$^{-1}$ to 
$-53$ kms$^{-1}$)]{\mbox{\resizebox{7.0cm}{!}{\includegraphics{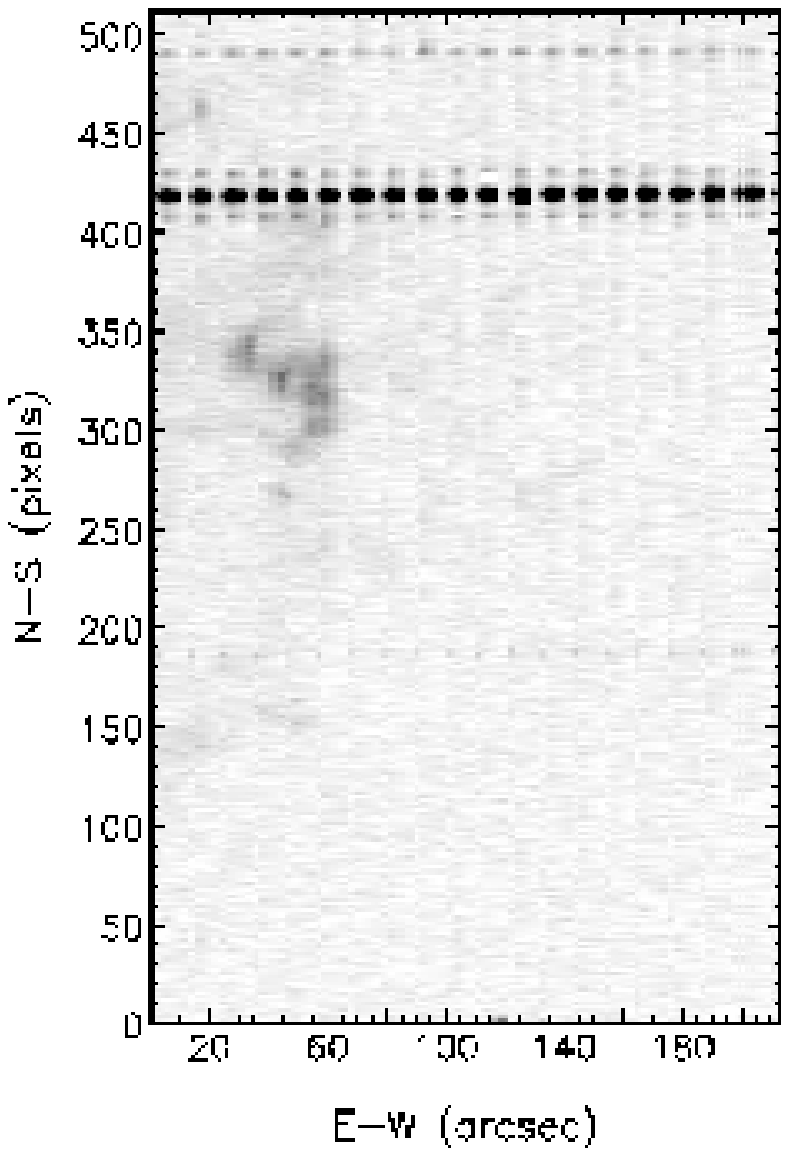}}}}
\caption{The knot shown at different velocity intervals, $I_{1}$, $I_{2}$ and 
$I_{3}$, (see Table 1). The images are shown at high-contrast in order to 
highlight substructure within the knot.}
\end{figure}
\end{landscape}

\begin{landscape}
\begin{figure}
\centering
\subfigure[$I_{1}$]{\mbox{\resizebox{7.0cm}{!}{\includegraphics{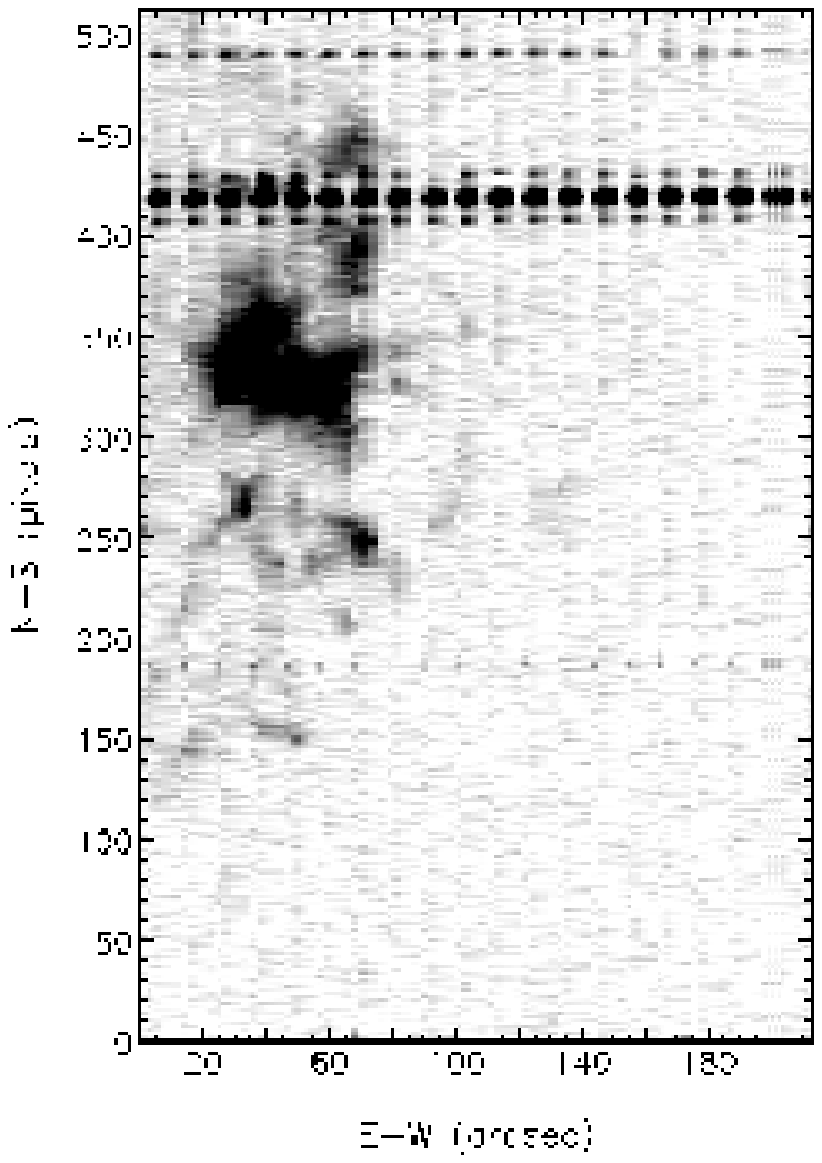}}}}
\subfigure[$I_{2}$]{\mbox{\resizebox{7.0cm}{!}{\includegraphics{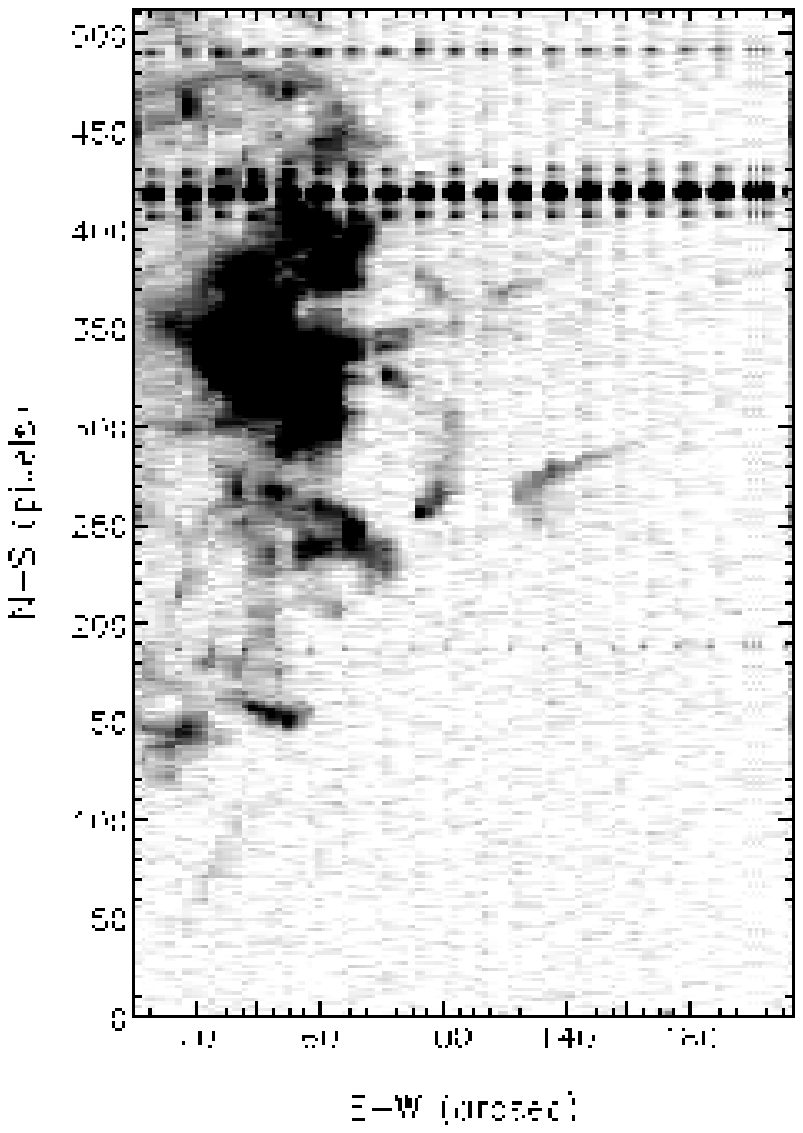}}}}
\subfigure[$I_{3}$]{\mbox{\resizebox{7.0cm}{!}{\includegraphics{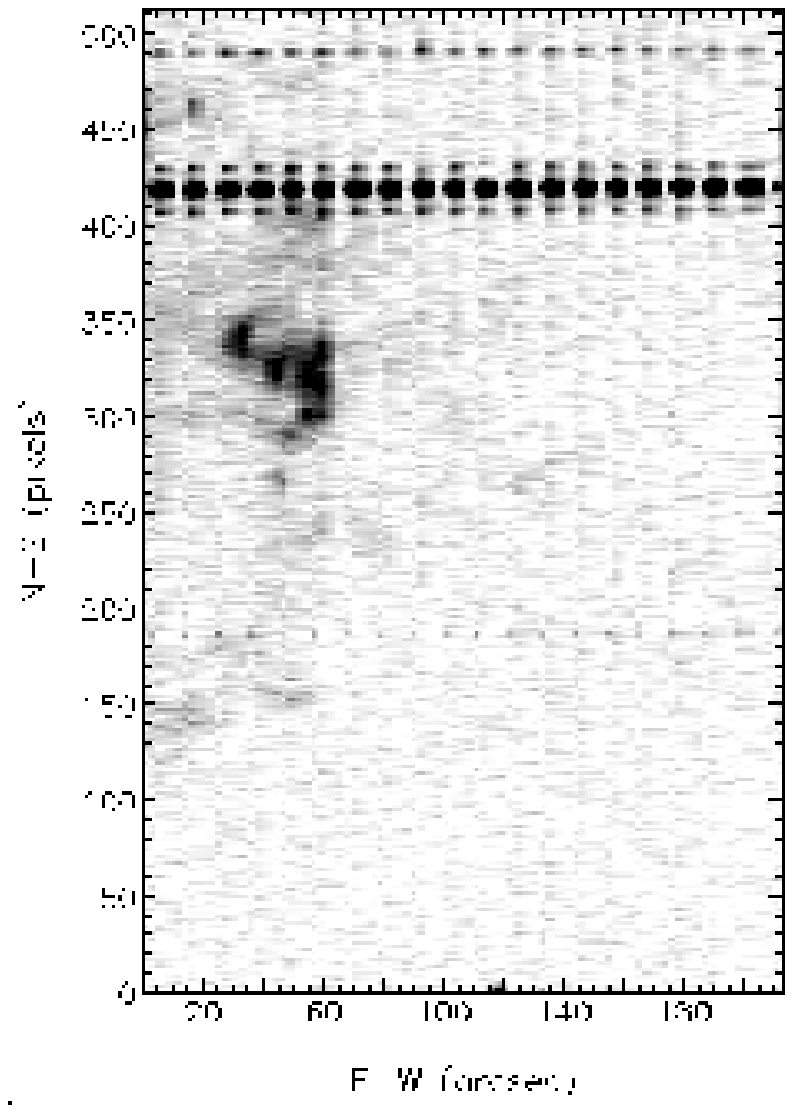}}}}
\caption{As Fig.~7, but shown at low-contrast in order to reveal faint 
substructure in the halo that surrounds the knot.}
\end{figure}
\end{landscape}

\begin{table} 
\centering
\begin{tabular}{ccc}
\hline
  Image    & $V_{min}$/kms$^{-1}$ & $V_{max}$/kms$^{-1}$\\ \hline
  $I_{1}$       & -83 & -73\\ 
  $I_{2}$       & -73 & -63\\ 
  $I_{3}$       & -63 & -53\\ \hline
  
\end{tabular}
\caption{Ranges in \vhel\ of the collapsed datacube 
slices that contain \OIII\ emission.}
\end{table}

(i)  Most of the \OIII\ emission from the filamentary halo surrounding the 
knot is present in $I_{2}$ (Fig.~8b). Halo material located to the south of 
the knot is also visible in $I_{1}$ (Fig.~8a), whereby the filaments situated 
at pixel 270, 30 arcsecs and pixel 250, 70 arcsecs are particularly bright in 
this image
. In contrast, the halo has virtually no red-shifted component of \vhel: 
$I_{3}$ (Fig.~8c) reveals very little diffuse emission, except for a compact 
feature at pixel 270, 50 arcsecs. The gas in the halo, therefore, has a very 
narrow
 velocity range. 

(ii) The knot gas has a large blue-shifted component of \vhel\ as similar 
levels of emission are present in $I_{1}$ and $I_{2}$. The knot gas 
surrounding the eastern 
half of the knot is particularly bright in $I_{1}$. The knot gas is completely
 absent in $I_{3}$. It appears that the knot gas has a more extensive velocity
 range than the faint halo gas.

(iii) Features A - K (Figs.~5 and 6) are all bright in $I_{2}$ (Fig.~8b). 
Trails H and I are both present in $I_{1}$, whereby trail H is particularly 
bright around pixel 360, 40 arcsecs (Fig~7a). The comet-like features, F and 
G, are also 
faintly present in $I_{1}$, as is the western half of plume G. All of these 
features are completely absent, however, in  $I_{3}$. 

(iv) Sub-knots A, B and C are much brighter than sub-knots E and F in $I_{1}$.
 This suggests that the western side of the knot has a higher recessional 
component of \vhel\ than the eastern side.

\subsection[]{Line profiles}

The manual fitting program \textsc{longslit} was used to fit
single Gaussian profiles to the \OIII\ line profiles from each slit. The 
profiles were first binned together into blocks
of two along the slit length that covered the bright knot and blocks of either
 three or four along the slit length that covered the surrounding, faint halo 
gas, in order to improve the signal-to-noise
ratio in each profile. The program was used
to produce plots of \vhel\ and the full width half maxima (FWHM) of the fitted profiles as a 
function of the slit length. The line profiles are
calibrated to $\pm$1 kms$^{-1}$ in absolute heliocentric velocity. The trends 
in \vhel\ from sections of four of the slits that cover the knot, $S_{7}$, 
$S_{11}$, $S_{13}$ and $S_{17}$, are shown in Figs.~9a - 9d (the positions of 
the slits are marked on Fig.~6). These slit positions were selected as they 
exhibit velocity trends that are representitive of those found at different 
spatial positions across the knot. The corresponding trends in FWHM of the fitted Gaussians are 
shown in Figs.~10a - 10d. The unit of cross-section along the y-axis in these 
figures corresponds to increments in pixels along the slit length. Note that 
as \vsys\ = $-68.5 \pm 1.0$ kms$^{-1}$, any \vhel\ greater than this value 
will be considered red-shifted and any \vhel\ less than this value will be 
considered blue-shifted. The velocity trends observed over each slit are 
summarised below:

\begin{figure*}
\centering
\mbox{\resizebox{\textwidth}{!}{\includegraphics{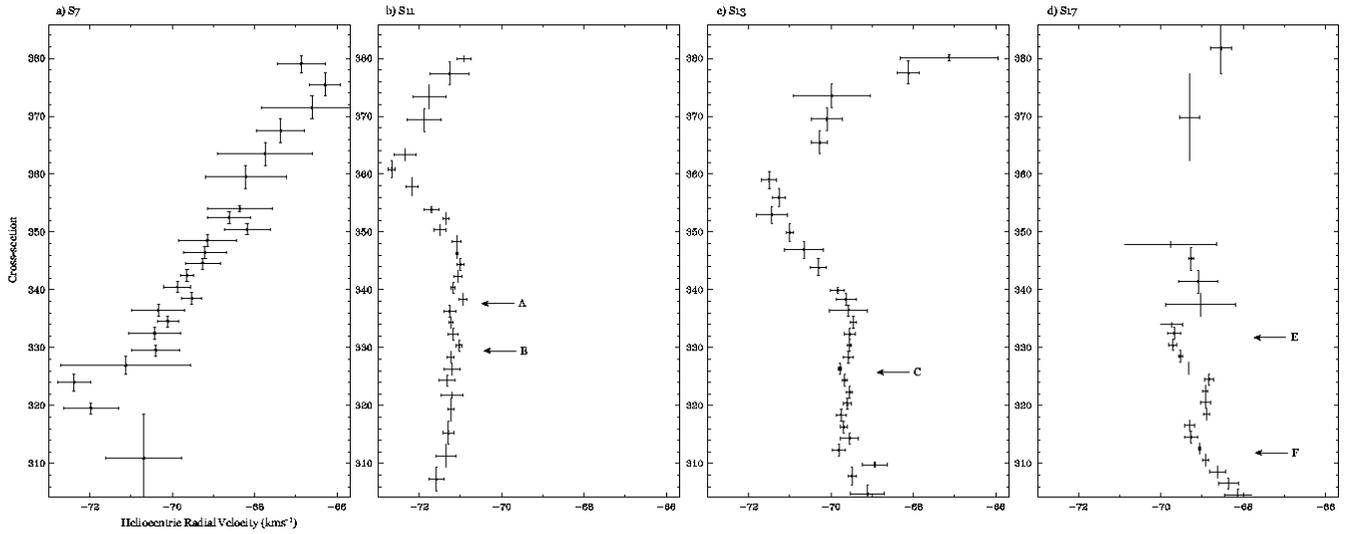}}}
\caption{Centroids of best-fit Gaussians to observed \OIII\ profiles 
($\equiv$ \vhel) from slit sections marked on Fig.~6. The positions of 
sub-knots A, B, C, E and F are labelled on the plots.}
\end{figure*}

\begin{figure*}
\centering
\mbox{\resizebox{\textwidth}{!}{\includegraphics{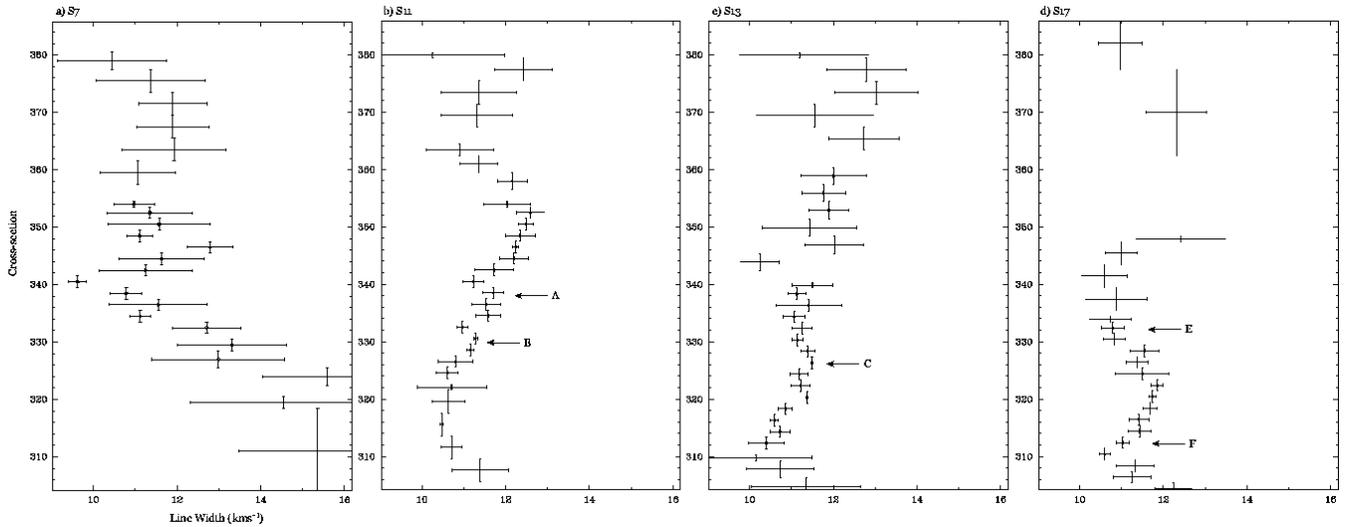}}}
\caption{Full width half maxima of best-fit Gaussians to observed \OIII\ 
profiles (see Fig.~9).}
\end{figure*}

(i) $S_{7}$: The gas bordering the eastern edge of the knot shows a trend of 
becoming increasingly red-shifted in a northwards direction (Fig~9a). The plot
 of line width against slit length (Fig.~10a) reveals a region of very inert 
gas at cross-section 340, which corresponds to the position of the eastern 
apex of the knot (Fig.~6). 

(ii) $S_{11}$: Sub-knots A and B have very similar values of \vhel, which lie 
close to \vsys\ of the knot (Fig.~9b). The knot gas immediately north of 
sub-knot A (cross-sections 346 - 362) shows increasingly blue-shifted 
velocities in a northwards direction. A reversal in velocity trend is observed
 at cross-section 365, whereby the Gaussian profile centres become 
increasingly red-shifted in a northwards direction. This reversal in velocity 
trend corresponds to the transition from the knot gas to the fainter gas 
identified as plume G in Fig.~6. The blue-shifted profiles observed in the 
knot gas region are broader than the profiles from the sub-knots and are 
generally broader than the red-shifted profiles observed in the region of the 
faint halo gas (Fig.~10b). The narrowest line profiles are associated with the
 faint halo gas to the south of the knot (cross-section 318). 

(iii) $S_{13}$: The plot of \vhel\ against slit length for $S_{13}$ (Fig.~9c) 
shows very similar trends to that of $S_{11}$ (Fig.~9b). The emission from 
sub-knot C (cross-sections 320 - 330) lies at \vsys\ of the knot. A trend 
towards blue-shifted velocities is observed in the knot gas region north of 
the knot between cross-sections 340 - 380. The observed line profiles from 
this blue-shifted knot gas are relatively broad (Fig.~10c) with an observed 
width of approximately 12 kms$^{-1}$. The faint halo gas located to the far 
north (cross-sections 380 - 400) of the field is red-shifted relative to \vsys\ of the knot. The line profiles from this faint halo gas are 
narrower than those from the knot gas. 

(iv) $S_{17}$: Sub-knots E and F correspond to the blue-shifted `dips' 
(cross-sections 330 and 315, respectively) in Fig.~9d. The sub-knots have 
narrower line profiles than the surrounding knot gas (Fig.~10d). The emission 
located south of sub-knot F (cross-sections 300 - 310) shows exceptional 
behaviour in that it is red-shifted relative to the sub-knots. Although this 
gas appears to be very close to the sub-knots and therefore we might expect it
 to constitute part of the knot gas, Fig. 5 shows that this red-shifted region
 of gas lies just outside of the knot-gas boundary. This suggests it may be a 
line-of-sight effect of a background filament in the halo.
 
\begin{figure}
\mbox{\resizebox{8.0cm}{!}{\includegraphics{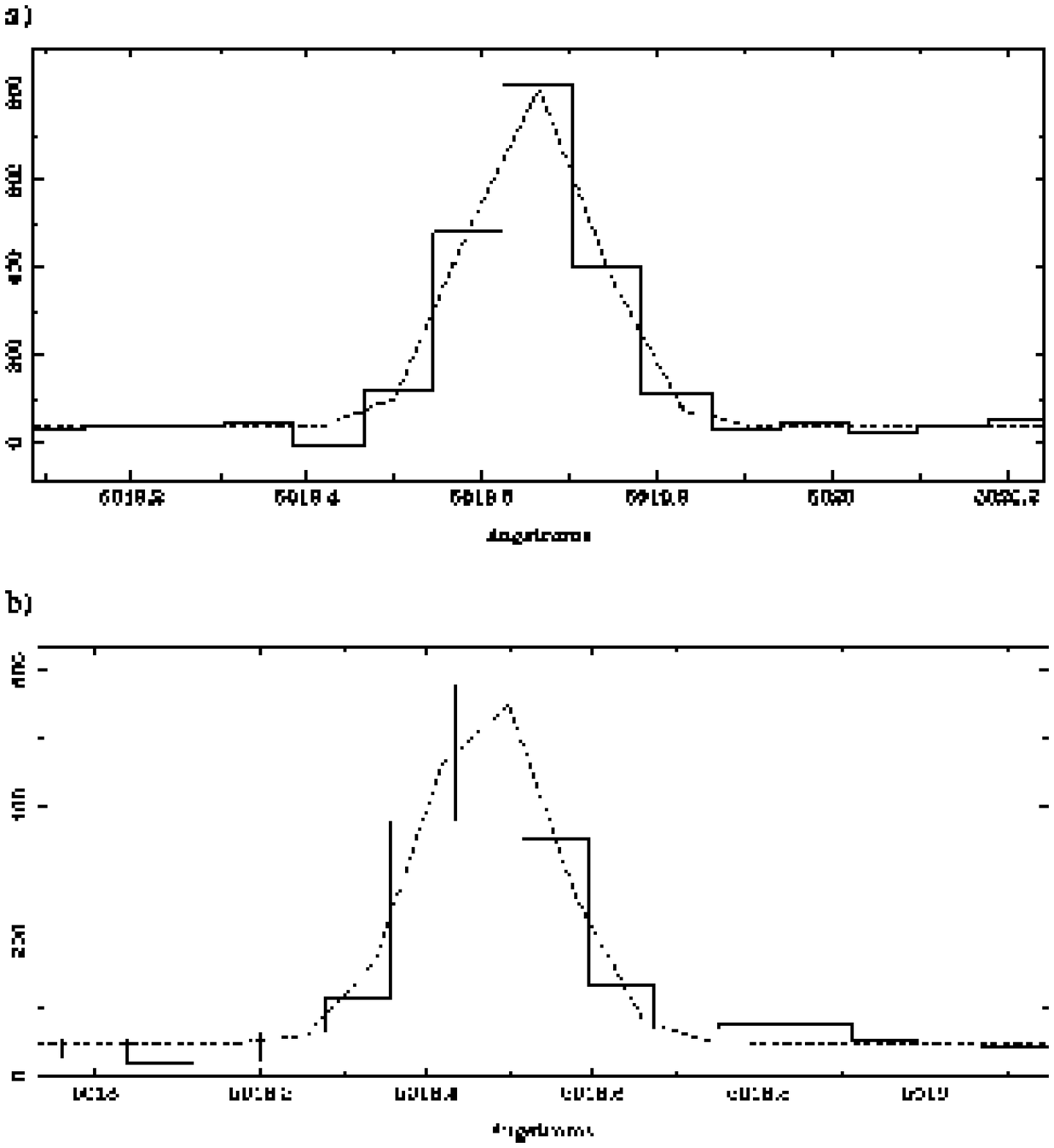}}}
\caption{Sample profiles showing single Gaussian fits to (a) a narrow profile from the knot gas at the eastern apex of the knot (S7 cross-sections 339 - 340), (b) a broad profile from the halo gas to the north of the knot (S11 cross-sections 379 - 381).}
\end{figure}

\section[]{Discussion}

The plots in Fig.~10 show that all of the observed \OIII\ profiles from the 
knot are very narrow (observed profile widths generally less than 12.5 
kms$^{-1}$). This supports the findings of \cite{1992MNRAS.254..477B}, who 
concluded that both this prominent knot and the whole halo are kinematically 
inert.

The width of an observed \OIII\ emission line profile is the result of the 
combination of several broadening mechanisms. In this case, the important 
broadening mechanisms are thermal broadening, instrumental broadening and 
turbulent broadening. The observed line profile can be modelled as a Gaussian 
profile of width $\Delta\mathrm{V}_{\mathrm{obs}}$, which is the convolution 
of the profiles corresponding to each of the individual broadening processes. 
If each individual component is also modelled as a Gaussian then 

\begin{equation}
\Delta\mathrm{V}_{\mathrm{obs}}^{2} = \Delta\mathrm{V}_{\mathrm{th}}^{2} + 
\Delta\mathrm{V}_{\mathrm{inst}}^2 + \Delta\mathrm{V}_{\mathrm{turb}}^{2} 
\end{equation}

where $\Delta\mathrm{V}_{\mathrm{th}}$, $\Delta\mathrm{V}_{\mathrm{inst}}$ and
 $\Delta\mathrm{V}_{\mathrm{turb}}$ are the values of the thermal broadening, 
instrumental broadening and turbulent broadening, respectively.

If the gas is in thermal equilibrium, then the thermal broadening is due to 
the Maxwellian distribution of the Doppler shifts of the emitted photons. This
 component has a full width at half maximum (FWHM) of 
 
\begin{equation}
\Delta\mathrm{V}_{\mathrm{th}} = 
2\sqrt{\frac{2\mathrm{k}\mathrm{T}_{\mathrm{e}}\mathrm{log}_{\mathrm{e}}2}{\mathrm{m}}} 
\end{equation}

where $\mathrm{k}$ is Boltzmann's constant, $\mathrm{T}_{\mathrm{e}}$ is the 
\OIII\ electron temperature and $\mathrm{m}$ is the atomic mass of oxygen.

The $\mathrm{T}_{\mathrm{e}}$ of 8860 K derived by \cite{1991MNRAS.252..535M} 
will be adopted (see Section 4.1) and this gives a 
$\Delta\mathrm{V}_{\mathrm{th}}$ of 5 kms$^{-1}$. The instrumental broadening 
can be determined from the FWHM of the Gaussians fitted to the arc line 
profiles, which in this case is approximately 7 kms$^{-1}$.

Sub-knots A to F all show similar velocity characteristics: each sub-knot has 
a $\Delta\mathrm{V}_{\mathrm{obs}}$ of approximately 11.5 kms$^{-1}$. Using 
Equation 1, this gives a typical $\Delta\mathrm{V}_{\mathrm{turb}}$ across the
 knot of 7.6 kms$^{-1}$. The sub-knots all exhibit a \vhel\ close to \vsys. 
The sub-knots are, therefore, very stable, kinematically inert objects.

The knot gas, however, exhibits complex motions over short spatial scales. 
Figs. 10a to 10d show that the line profiles of the knot gas (cross-sections  350 to 360 and 300 to 310) 
are generally broader than the line profiles of the sub-knots, with a 
$\Delta\mathrm{V}_{\mathrm{turb}}$ of approximately 9.0 kms$^{-1}$ [note that 
these profiles are only broad relative to the profiles of the sub-knots and 
are still much narrower than the typical line widths observed from the bright 
core of NGC~6543 (Meaburn et al. 1991)]. Both the velocity slices (Figs. 7 and
 8) and the plots of \vhel\ (Fig. 9) show that knot gas has a larger 
blue-shifted component of \vhel\ than both the sub-knots and the halo gas. This is most likely an optical depth effect in the dense knot; 
only the expansion from the near-side of the globule is seen and the
far-side emission is absorbed.

In contrast, the observed line widths of the knot gas surrounding the eastern 
apex of the knot (cross-section 340) are very narrow (Fig. 10a). Using 
Equation 1 it is found that the gas in this region has a 
$\Delta\mathrm{V}_{\mathrm{turb}}$ of just 4.7 kms$^{-1}$. The sample profiles shown in Fig. 11 demonstrate that it
is appropriate to fit both the narrow and broad profiles
with single Gaussians albeit with different widths.

The bright knot gas and the surrounding faint halo differ in their velocity 
characteristics. The line profiles in the faint halo gas are generally narrower
 than in the knot gas. In addition, there is a trend towards increasingly 
red-shifted velocities moving northwards from the boundary between the two 
regions (Figs.~9a - c). 

The relatively turbulent nature of the knot gas surrounding the sub-knots 
provides evidence of a gas flow around the knot. The very narrow line profiles
 in the east are most likely to coincide with the apex of the flow. The relatively broad, blue-shifted profiles 
associated with trail H (Fig. 9b cross-section 360) indicate that the trail is
 most likely to be material that has been ablated from the knot rather than a 
filament in the background halo. 

The gas kinematics reported here are most likely to be the result of photo-ionisation of the surface a pre-existing dense clump followed by localised expansion of the ionised gas. The ionised gas is overpressured with respect to the surroundings and expands outwards from those portions of the clump facing the star. The portions of the clump not exposed to the direct ionizing flux of the star are in shadow but will experience a weaker diffuse radiation field that is typically $\sim 15\%$ of the local direct field (Canto et al. 1998). The shadow region behind a neutral clump is cooler and underpressured compared to the surrounding ionized gas and is thus compressed. If this higher density gas is ionised by the diffuse flux then it can appear as a tail behind the clump (Canto et al 1998). Such a mechanism seems plausible for generating the tail structures seen around the knot and sub-knots in NGC~6543. The radiation pressure is too feeble at these distances to cause any effect on the knot while, as argued below, the wind from the star does not reach to the distance of the knots.

The adiabatic sound speed of the gas, 
$\mathrm{c}_{\mathrm{i}}$, as it evaporates off the surface of the knot is 
given by
\begin{equation}
\mathrm{c}_{\mathrm{i}} =
 \sqrt{\frac{\gamma\mathrm{k}\mathrm{T}_{\mathrm{e}}}{\mu\mathrm{m}_{\mathrm{H}}}} 
\end{equation}
where $\gamma$ is the specific heat ratio of a monatomic gas (= 5/3), 
$\mathrm{T}_{\mathrm{e}}$ is the electron temperature of the gas, $\mu$ is the
 mean atomic weight of ionised hydrogen (= 0.5) and $\mathrm{m}_{\mathrm{H}}$ 
is the mass of a hydrogen atom (= 1.7 $\times 10^{-27}$kg). Assuming that the 
ionised gas is at the temperature derived by \cite{1991MNRAS.252..535M} of 
8800K, equation 3 gives $\mathrm{C}_{\mathrm{i}}$ = 16.5 kms$^{-1}$.  
 
The flows observed in the vicinity of the sub-knots exhibit 
$\Delta\mathrm{V}_{\mathrm{turb}}$ less than 15 kms$^{-1}$. This is consistent
 with the sound speed expected as gas flows down the pressure gradient as it 
evaporates off an ionised globule surface.
 
\subsection[]{Knot formation and evolution}

Inhomogenities are ubiquitous in PNe (e.g. O'Dell et al. 2002)
ranging from compact cometary knots (e.g. the Helix and Eskimo PNe)
to large--scale, lower density clumps in the extended halos such
as those being considered here. 

In general, \cite{1994ApJ...428..186V} predicted that the 
postshocked gas at the interface of the fast and slow winds in a PN will 
become hydrodynamically unstable and non-linear instabilities will develop. 
\cite{1989MNRAS.241..625D} argued that the compact cometary knots 
in the Helix nebula represent surviving 
condensations from the atmosphere of the progenitor Red Giant star, 
which are ejected with the halo during the slow wind phase. There is now evidence for clumpy shell structure in AGB stars in advanced evolutionary stages: K'-band imaging of the AGB star IRC+10216 by \cite{1998A&A...333L..51W} revealed its dusty shell is highly fragmented. They predict that the knots are formed by fragmentation of the circumstellar shell due to large-scale surface convection cells. \cite{1973ApJ...179..495C} suggested that the 
knots in the halo of NGC 6543 (Fig. 1) are formed far from 
the central star as a result of Rayleigh-Taylor 
instabilities forming where the ionisation front expands into the neutral 
Red Giant wind ejecta. 

\cite{2004ApJ...612..319S} describe the two possible mechanisms responsible 
for the subsequent shaping of clumps and filaments after they have formed within PNe: (1) photoevaporation by 
ionising radiation from the central star; (2) interaction with a fast stellar 
wind. There are thus two possiblities for the formation and subsequent 
evolution for the knot in NGC~6543, both of which will be assessed based on 
previous evidence and the new results presented in this paper. Since the flow velocities reported here are clearly subsonic with no evidence of any supersonic motions generated by a fast wind, photoevaporation (Case 1) must dominate the structure and dynamics. Previously \cite{1992MNRAS.254..477B} measured the global expansion
 of the whole halo as $\leq$ 10 kms$^{-1}$, which is consistent with the 
expected expansion of a Red Giant wind (Corradi et al. 2003). Clearly we are considering
an earlier stage in the mass loss history of the central star than
the post--AGB phase and the onset of the fast wind.

Until recently, the presence of the fast stellar wind in the halo was 
considered necessary in order to account for the discrepency between the 
$\mathrm{T}_{\mathrm{e}}$ derived by \cite{1989MNRAS.239....1M} and 
\cite{1991MNRAS.252..535M}, which was originally explained using the 
mass-loaded wind model (see section 1). However, an alternative explanation was proposed by 
\cite{1994MNRAS.271..993V}. They were able to show that if clumps with a 
density greater than 10$^{6}$ cm$^{-3}$ are present in a PN, then the \OIII\ 
nebular line is collisionally de-excited instead of undergoing a forbidden 
transition, and hence the average [O~{\sc iii}]$\lambda\lambda$4959 + 
5007/4363 line ratio would be higher and the derived $\mathrm{T}_{\mathrm{e}}$
 overestimated.  Clumps in PNe have been found to have central densities of 
$\sim$ 10$^{6}$ cm$^{-3}$ (Meaburn et al. 1992, Borkowski et al. 1993, 
Dopita et al. 1994, Bautista, Pradhan \& Osterbrock 1994). It is reasonable to
 assume, therefore, that the knot in NGC~6543 has a density of the same order 
of magnitude.

In summary, it appears that the fast wind, or even a mass--loaded flow driven by the fast wind, is not impinging on this halo knot in NGC~6543. The fast wind remains contained in 
the central pressure driven bubble in the bright AGB superwind that forms the 
nebular core.  The inner shock front will stop the fast wind having any effect
 on the outer halo at the present time. This conclusion rules out the possibility of the knottiness of the halo being due to instabilities between the fast wind and the earlier ejected
 Red Giant wind since the former has not yet reached the latter. 
Furthermore, knots formed via Rayleigh-Taylor instabilities are predicted to 
have radially symmetric structures (O'Dell et al. 2002), which is not observed
 in this case. 

\subsection{Knot ionization}

The ionizing flux of the central star, $S_*$, can be estimated from the $\rm H\beta$ flux of $2\times 10^{-10}~{\rm erg~cm^{-2}~s^{-1}}$ measured by \cite{1986A&A...169..227B} using the method described in Osterbrock (1989). This gives $S_*=2\times 10^{45}~s^{-1}$. Using the distance of 1Kpc (Reed et al. 1999) gives that that knot lies at $2.5\times 10^{18}~{\rm cm}$ from the star and so the flux incident on the knot at this distance is $F_*={S_*/{4\pi R^2}}=3\times 10^{8}~{\rm s^{-1}~cm^{-2}}$.

The ionization balance in a photoevaporating knot is given approximately by (L\'{o}pez-Mart\'{i}n~et~al 2001; Henney 2001) 
\begin{equation}
F_*=un_i + \alpha n_i^2 h
\end{equation}
where $u$ is the expansion of newly ionized material away from the ionization front, $n_i$ is the density of the ionized gas and $h\sim 0.1 r_{\rm clump}$ is the thickness of the ionized layer around the knot. The first term on the right hand side of equation 4 is due to the ionization of fresh neutral material while the second term is due to reionizing recombined material. For a given local ionizing flux, there is a critical density in the ionized gas that determines which process dominates the absorbtion of ionizing photons. This is important because, as \cite{2001ApJ...548..288L} point out, if recombinations dominate then the surface brightness of an ionized knot matches the incident ionizing flux. If recombinations are insignificant, then the emission from the clump is much reduced and the clump will appear fainter than expected from photon balance. \cite{2001RMxAC..10...57H} shows that most photoevaporated flows (proplyds, Eagle pillars, Rosette globules) are recombination dominated unless the ionizing flux is small or if the distance of the clump from the source is large (Gum globules, Helix knots). 

The above arguments suggest a very straightforward explanation for the puzzling brightness difference between the bright knot and other halo structures in NGC 6543 (see Fig 1). The knot the is densest part of the halo and widespread recombinations lead to a large surface brightness; essentially all of the incident ionizing photons are absorbed and re-emitted. Meanwhile, the lower density of the other features in the halo mean that they are already fully ionized or recombinations are insignificant. Using the flux value above,  $F_*=3\times 10^{8}~{\rm s^{-1}~cm^{-2}}$ and an ionized layer of $1.5\times 10^{16}~{\rm cm}$ then from Figure 1 of Henney (2001), we require that the ionized gas around the knot have a density $>> 3\times 10^2~{\rm cm^{-3}}$ while the other halo features must be less dense than this. It would be interesting to use the [S~{\sc ii}] doublet to   measure the local electron density in the knot and the other halo features.

\section[]{Conclusions}

The results presented in this paper reveal for the first time the existence of
 a low-turbulence, low-velocity gas flow around the most prominent knot in the
 halo of NGC~6543. The flow velocities observed are comparable to the sound 
speed as gas flows down the pressure gradient as it photoevaporates off an 
ionised globule surface. Although it is known that the fast stellar wind has 
played an important role in shaping the complex core of the Cat's Eye nebula 
(Balick \& Preston 1987), we find it unlikely that it has percolated into the 
halo, as the observed flow velocities are too low to be the result of an 
interaction of the fast stellar wind with the knot. We conclude that
 the fast wind must still be contained in a central pressure-driven bubble. 
Consequently, the knot cannot have been formed via an instability as the fast 
wind interacts with the slower Red Giant wind, and it is thus likely that the knot is
 a relic of the Red Giant wind. We suggest that the knot is brighter than other halo structures because it is the only feature dense enough for the ionizing photon flux to be balanced by recombinations behind the knot ionization front.  

The kinematical analysis of the knot in NGC~6543 should aid understanding of 
how these structures form and evolve in general. A natural follow-up to these 
observations would be a detailed analysis of the chemical abundances of the 
knot and the dynamical evolution of the sub-knots (A - F) at high spatial 
resolution.

\section*{Acknowledgments}

DLM thanks PPARC for her research studentship. We would like to thank the 
staff at the San Pedro Martir observatory who helped with the observations. 
Thanks also to Malcolm Currie at Starlink who wrote the program 
\textsc{formcube} and 
Romano Corradi for allowing us to use his composite image of the halo of 
NGC~6543 (Fig. 1).

\label{lastpage}

\end{document}